%% file: main.tex
\documentclass[12pt,a4paper]{amsart}
\usepackage[colorlinks,citecolor=blue,urlcolor=blue,bookmarks=false,hypertexnames=true]{hyperref}
\include{preamble}
\usepackage{geometry}
\title[Locally Conformal higher order  Lagrangian Dynamics]{Locally Conformal higher order  Lagrangian Dynamics}

\author{Serdar Ç\.ite}
\address{Department of Physics, Boğaziçi University, Istanbul, Turkey   }
\email{serdar.cite@std.bogazici.edu.tr, Corresponding author}

\author{O\u{g}ul  Esen}
\address{Department of Mathematics, Gebze Technical University,    41400 Gebze-Kocaeli, Turkey,
\newline 
Center for Mathematics and its Applications, Khazar University, AZ1096 Baku, Azerbaijan}
\email{oesen@gtu.edu.tr}



\begin{document}

\begin{abstract} This work presents higher order Lagrangian dynamics possessing locally conformal character. More concretely, locally conformal higher order Euler-Lagrange equations are written with particular focus on the second- and the third-order cases.  
\end{abstract}

\maketitle

\tableofcontents

\doublespacing

 \section{Introduction}

The existence of global Hamiltonian dynamics defined on a whole (symplectic or Poisson) manifold reads local Hamiltonian dynamics in each local chart, 
\cite{AbMa78,Arnold-book,LiMa-book}.
However, the inverse of this assertion does not generally hold; having local Hamiltonian dynamics in each chart does not guarantee the existence of global Hamiltonian dynamics. 
An example of this is provided by locally conformally symplectic (abbriviated as LCS) manifolds and the dynamics on this manifold that is locally conformal Hamiltonian dynamics \cite{Banyaga,LeeHC,Stanciu2019,vaisman, WojtkowskiLiverani1998}. 


Locally conformal analysis has been extensively studied through various approaches in recent literature. We refer to \cite{ChantraineMurphy2019,OtSt15} for foundational work on LCS geometry in the context of cotangent bundles, which underpins the rationale of the present work, given that cotangent bundles can be considered generic for LCS manifolds. In \cite{Stanciu2019}, reduction theories have been applied to LCS dynamics. For a discussion on Hamilton-Jacobi theory within an LCS framework, see \cite{EsenLeonSarZaj1}. In the context of field theories, see \cite{EsLeSaZa21,EsLeSaZa21b}, and for time-dependent Hamiltonian dynamics, refer to \cite{AtEsLeSa23}.

An obvious question is to investigate the Lagrangian counterpart of locally conformal dynamics. This is established in \cite{MR4747740,esen2024variational} by computing locally conformal Euler-Lagrange equations. 

In the present work, we extend locally conformal Lagrangian dynamics to higher order tangent bundles. This includes dynamical equations governed by Lagrangian functions that depend on higher order terms beyond just position and velocity. More concretely, we present the locally conformal higher order Euler-Lagrange equations. 

In Section \ref{sec:lcs}, we provide a concise yet explicit presentation of locally conformal Hamiltonian dynamics, whereas in Section \ref{LCEL-sec}, we review locally conformal Lagrangian dynamics. Section \ref{Sec-Cont} summarizes the fundamentals of higher order Lagrangian theory. Before proceeding to the $n$-th order theories, we derive our first novel result—the locally conformal second order Euler-Lagrange equations—in Section \ref{sec:2ndLC}. In Section \ref{sec:-Bell}, we examine locally conformal $n$-th order Lagrangian functions, while Section \ref{sec:-Bell-} introduces a compact notation for the terms required in $n$-th order Lagrangian dynamics within the locally conformal framework. Subsequently, in Section \ref{sec:LCnEL}, we present our novel result: the locally conformal $n$-th order Euler-Lagrange equations. Finally, Section \ref{sec:examp} provides an example that applies the theories we have developed—a locally conformal analysis of the chiral oscillator.

\section{Locally Conformally Symplectic Geometry}\label{sec:lcs}

There are alternative ways to define locally conformally symplectic (LCS) manifolds. We begin with the one involving the Lichnerowicz-deRham (LdR) differential then we shall elaborate what we call the local-to-global approach. 

\bigskip
\noindent 
\textbf{Locally Conformally Symplectic Manifolds.} An LCS manifold $(M,\Omega,\varphi)$ is consisting of an even dimensional manifold $M$ and an almost symplectic two-form $\Omega$ (non-degenerate but not closed) satisfying 
\begin{equation}\label{eqw}
d_\varphi\Omega:=d\Omega-\varphi\wedge \Omega=0
\end{equation}
for a closed (called as Lee)  one-form $\varphi$, 
\cite{Bazzoni2018,LeeHC,Libermann-LCS,vaisman}. 
Here, $d_\varphi$ stands for the Lichnerowicz-deRham differential. For a Hamiltonian function $H$, the locally conformal Hamiltonian dynamics is determined through the locally conformal Hamilton's equation:
\begin{equation}\label{semiglobal}
   \iota_{\xi_{H}}\Omega=d_\varphi H,
\end{equation}
where $\xi_H$ called locally conformal Hamiltonian vector field. 

If, in \eqref{eqw}, the Lee-form is identically zero then $\Omega$ turns out to be a symplectic two-form hence we arrive at a symplectic manifold $(M,\Omega)$. In this case, the locally conformal dynamics in \eqref{semiglobal} becomes the Hamiltonian equation in its very classical form:
\begin{equation}\label{semiglobal+}
   \iota_{X_{H}}\Omega=d H. 
\end{equation}

The generic example of a symplectic manifold is the cotangent bundle $T^*Q$ of a manifold $Q$ \cite{AbMa78,LiMa-book,MaRa99}. In this case, the canonical symplectic two-form on $T^*Q$ is given by the negative exterior derivative 
\begin{equation}
\Omega_Q=-d\Theta_Q
\end{equation}
of the canonical (Liouville) one-form $\Theta_{Q}$. 

Generic examples of LCS manifolds are cotangent bundles as well \cite{ChantraineMurphy2019}. 
 Consider a closed one-form $\vartheta$ on a manifold $Q$. Pull this one-form back to the cotangent bundle using the cotangent bundle projection $\pi_Q:T^*Q\mapsto Q$, then obtain a closed and semi-basic one-form 
\begin{equation}\label{varphi}
    \varphi = (\pi_Q)^* \vartheta.
\end{equation}
The negative LdR differential with respect to $\varphi$ of the canonical one-form $\Theta_Q$ is an LCS two-form
\begin{equation}\label{Omega-phi}
\Omega_\varphi=-d_\varphi \Theta_Q = - d\Theta_Q + \varphi\wedge \Theta_Q  = \Omega_Q + \varphi\wedge \Theta_Q .
\end{equation} 
Therefore, the three-tuple  $
(T^*Q,\Omega_\varphi,\varphi)$  is an LCS manifold. 

\bigskip
\noindent 
\textbf{Local to Global Analysis.} For an LCS manifold $(M,\Omega,\varphi)$, each local chart of the manifold $M$ is a symplectic space $(U_\alpha,\Omega_\alpha)$ where $\Omega_\alpha$ is a local symplectic two-form. 
On the intersection $U_\alpha \cap U_\beta$ of two local symplectic charts $(U_\alpha,\Omega_\alpha)$ and $(U_\beta,\Omega_\beta)$ of $M$, we have the equality 
\begin{equation}\label{sympl-rel-}
   e^{\sigma_\alpha}\Omega_\alpha = e^{\sigma_\beta}\Omega_\beta,
\end{equation}
where $\{e^{\sigma_\alpha}\}$'s are called conformal factors. 
The existences of the conformal factors read that local symplectic two-forms $\{\Omega_\alpha\}$ cannot be  glued up to a (real-valued) two-form on the whole manifold $M$. On the other hand, we can define local two-forms 
\begin{equation}
\Omega\vert_\alpha=e^{\sigma_\alpha}\Omega_\alpha,
\end{equation}
which fail to be symplectic since they are not closed. But one can glue up local two-forms $\{\Omega\vert_\alpha\}$ define a global two-form which we denote by $\Omega$. This two-form is the one in the definition of the LCS manifold $(M,\Omega,\varphi)$. 
The exterior derivative 
\begin{equation}\label{hola}
 d\Omega\vert_\alpha=d\sigma_\alpha\wedge \Omega\vert_\alpha,
\end{equation}
determines the Lee one-form $\varphi\vert_\alpha=d\sigma_\alpha$ in the local picture.

For the dynamics, we start with the local symplectic Hamiltonian flow on a local symplectic space $(U_\alpha,\Omega_\alpha)$ given by 
\begin{equation}\label{geohamalpha}
\iota_{X_{\alpha}}\Omega_{\alpha}=dH_{\alpha},
\end{equation}
where $X_{\alpha}$ is the local Hamiltonian vector field. On the intersection $U_\alpha\cap U_\beta$, we assume the local equality 
\begin{equation}\label{Fatih}
e^{\sigma_\alpha}H_\alpha=e^{\sigma_\beta}H_\beta.
\end{equation}
This leads us to determine a global function $H$ on $M$ so that its local picture is  
\begin{equation} \label{glueHamFunc}
H\vert_\alpha=e^{\sigma_\alpha}H_\alpha.
\end{equation}
In accordance with, we glue the local Hamilton's equation \eqref{geohamalpha} to the LCS Hamilton's equation \eqref{semiglobal}.

\bigskip
\noindent 
\textbf{Coordinate Realizations.} Consider the cotangent bundle  LCS manifold given by $
(T^*Q,\Omega_\varphi,\varphi)$. Assume an open covering for a manifold $Q$ with dimension $r$: 
\begin{equation}\label{chart}
Q=\bigsqcup_\alpha V_\alpha . 
\end{equation}
This induces manifold structures on the tangent and the cotangent bundles as
\begin{equation}\label{chartsTQ}
TQ=\bigsqcup_\alpha TV_\alpha = \bigsqcup_\alpha V_\alpha \times \mathbb{R}^r, \qquad T^*Q= \bigsqcup_\alpha T^*V_\alpha = \bigsqcup_\alpha V_\alpha \times (\mathbb{R}^r)^*,
\end{equation}
respectively. The Darboux coordinates $(q^i,p_i)$ on $T^*V_\alpha$ reads the global canonical one-form $\Theta_{Q}$ and the global symplectic two-form $\Omega_Q$ as
\begin{equation}\label{bar-bar}
\Theta_{Q}\vert_\alpha=p_idq^i, \qquad \Omega_Q\vert_\alpha=dq^i\wedge dp_i,
\end{equation}
respectively. On the other hand, there exist Darboux coordinates $(q^i,r_i)$ that yield the local canonical one-form $\Theta_\alpha$ (which is not global) and the local canonical symplectic two-form $\Omega_\alpha$ (which is also not global) as
\begin{equation}\label{r-p}
\Theta_\alpha = r_i dq^i, \qquad \Omega_\alpha=dq^i \wedge dr_i,
\end{equation}
respectively. The relationship between local momentum coordinates is 
\begin{equation}
r_i = e^{-\sigma_\alpha}p_i. 
\end{equation}
Without loss of generalization, we take the conformal factors $\sigma_\alpha=\sigma_\alpha(q)$ depend only on the coordinates of the base manifold $Q$. 
Then, the local symplectic two-form $\Omega_\alpha$ in \eqref{bar-bar}, in terms of the coordinates $(q^i,p_i)$, is computed to be
\begin{equation}
\Omega_\alpha= -d\Theta_\alpha = dq^i \wedge dr_i= e^{-\sigma_\alpha} \Big (dq^i\wedge dp_i + \frac{1}{2} A_{ij}dq^i\wedge dq^j \big), 
\end{equation}
where we define the following tensorial object: 
\begin{equation}\label{Aij}
A_{ij}:= \varphi_ip_j -\varphi_jp_i, \qquad \varphi_i:=\frac{\partial \sigma_\alpha}{\partial q^i}.
\end{equation}
Thus LCS two-form $\Omega_\varphi$ defined in \eqref{Omega-phi} is written in terms of the coordinates $(q^i,p_i)$ as
\begin{equation}\label{twist}
\Omega_\varphi\vert_\alpha = dq^i\wedge dp_i + \frac{1}{2} A_{ij}dq^i\wedge dq^j .
\end{equation}
In terms of the Darboux coordinates $(q^i, p_i)$, writing $H$ instead of $H_\alpha$, we arrive at the following proposition which states the Hamiltonian motion possessing local conformal character.
\begin{proposition}
In Darboux' coordinates $(q^i,p_i)$, the LCS Hamilton's equation generated by a Hamiltonian function $H=H(q,p)$ is
\begin{equation}\label{LCSHE}
\frac{dq^i}{dt} = \frac{\partial H}{\partial p_i}, \qquad
\frac{dp_i}{dt} = -\frac{\partial H}{\partial q^i} - A_{ij} \frac{\partial H}{\partial p_j} + H\varphi_i,
\end{equation}    
where the quantities $A_{ij}$ and $\varphi_i$ are those given in \eqref{Aij}.
\end{proposition}

\section{Locally Conformal Lagrangian Dynamics}\label{LCEL-sec} 

 Consider the tangent bundle $TQ$ covered by local charts $TV_\alpha$ (as defined in \eqref{chartsTQ}), each equipped with a local Lagrangian function $L_\alpha=L_\alpha(q^i,\dot{q}^i)$. It is assumed that these local Lagrangian functions satisfy conformal relations on intersections, say $TV_\alpha \cap TV_\beta$, as
\begin{equation}\label{laglog}
e^{\sigma_\alpha(q)}L_\alpha (q,\dot{q}) =e^{\sigma_\beta(q)}L_\beta (q,\dot{q}) ,
\end{equation}
where $\{\sigma_\alpha\}$'s represent a family of smooth functions. Once again, the conformal factor does not permit us to glue the local Lagrangian functions $\{L_\alpha\}$ into a   real-valued Lagrangian function on the tangent bundle $TQ$. However, local functions
\begin{equation}\label{loc-Lagrange}
L\vert_\alpha = e^{\sigma_\alpha(q)}L_\alpha.
\end{equation}
determine a global real-valued function $L$ on $TQ$. Notice that, without loss of generality, we choose the conformal factor to depend only on the position $(q^i)$. See the discussion done in the previous section, or \cite{esen2024variational} for the reasoning behind this choice.

      In the induced coordinates $(q^i,\dot{q}^i)$ on the tangent bundle $TQ$,  we arrive at the following proposition  \cite{esen2024variational,MR4747740} writing by using the notation presented in this subsection.
\begin{proposition}\label{prop-1}
    For a given Lagrangian function $L=L(q,\dot{q})$, the locally conformal Euler-Lagrange equations (of the first order) is      
         \begin{equation} \label{LCEL}
   \frac{\partial L }{\partial q^i} - \frac{d }{dt} \frac{\partial L }{\partial \dot{q}^i}  =  \mathcal{A}_i [L] 
,
   \end{equation} 
   where the quantity on the right hand side is 
    \begin{equation} \label{A-1}
   \mathcal{A}_i[L] =\varphi_i L -\varphi_j \dot{q}^j  \frac{\partial L }{\partial \dot{q}^i} .
   \end{equation} 
\end{proposition}
The locally conformal Hamilton's equations in \eqref{LCSHE} and the locally conformal Euler-Lagrange equations in \eqref{A-1} are related with the Legendre transformation:
\begin{equation}\label{local-FL}
\mathbb{F}L_\alpha: TV_\alpha \longrightarrow T^*V_\alpha, \qquad (q^i,\dot{q}^i) \mapsto (q^i,r_i)=\big(q^i,\frac{\partial L_\alpha}{\partial \dot{q}^i}\big).
\end{equation}

See that when the conformal factor is zero, we have $\varphi_\alpha=0$ hence $\mathcal{A}_i=0$ then from  the locally conformal Euler-Lagrange equations (of the first order) \eqref{LCEL}, we obtain the Euler-Lagrange equation in its very classical form:
\begin{equation}\label{EL}
\frac{\partial L }{\partial q^i} - \frac{d }{dt} \frac{\partial L }{\partial \dot{q}^i}  =0.
\end{equation}

\section{Higher Order Lagrangian Dynamics} \label{Sec-Cont}
In the classical picture \eqref{EL}, the Lagrangian function is defined on the tangent bundle $TQ$  hence it depends on position and velocity, that is, $L=L(q,\dot{q})$. The variation of the action integral yields the Euler-Lagrange equations \eqref{EL}. To generalize this, we may define higher order Lagrangian functions on the higher order tangent bundles,  \cite{AbMa78,MR65391,CrSaCa86}. In this subsection, we briefly recall the higher order Euler-Lagrange equations, with particular interest in the second order theory.

We denote the $n$-th order tangent bundle by $T^nQ$ with induced coordinates  $(q^i,\dot{q}^i,\ddot{q}^i, \dots, q^i_{(n)})$. In this case, the Lagrangian 
$L$ depends on the base coordinates and their derivatives up to the $n$-th order, that is, 
$L=L(q,\dot{q},\ddot{q}, \dots, q_{(n)})$, where $q^i_{(n)}$ denotes the $n$-th time derivative of the position $q^i$. Consequently, the variation of the action integral yields the higher order Euler-Lagrange equations, given by 
\begin{equation}\label{HOEL+}
 \sum_{s=0}^{n} (-1)^s \frac{d^s}{dt^s}   \frac{\partial L  }{\partial {q}^i_{(s)} }       =0.
\end{equation}

 A second order Lagrangian function is defined on the second order tangent bundle  $T^2Q$, so it depends on the acceleration in addition to position and velocity. That is,  
$L=L(q,\dot{q},\ddot{q})$. If $Q$ is an $r$-dimensional manifold then $T^2Q$ is a $3r$ dimensional. For a second order Lagrangian function, we have the second order Euler-Lagrange equations as
\begin{equation}\label{2EL}
\frac{\partial L }{\partial q^i} - \frac{d }{dt} \frac{\partial L }{\partial \dot{q}^i}  +  \frac{d^2 }{dt^2} \frac{\partial L }{\partial \ddot{q}^i}  =0.
\end{equation}

\section{Locally Conformal Second Order Lagrangian Dynamics} \label{sec:2ndLC}

We begin with the locally conformal analysis of the second order Lagrangian dynamics \eqref{2EL} on $T^2Q$. Consider the local charts $\{V_\alpha\}$, as given in \eqref{chart}, for $Q$. We then define the local chart for the second order tangent bundle as 
\begin{equation}
 T^2Q=\bigsqcup_\alpha T^2V_\alpha = \bigsqcup_\alpha V_\alpha \times \mathbb{R}^r \times \mathbb{R}^r. 
\end{equation}
We consider a local Lagrangian on $T^2V_\alpha$ as
\begin{equation}
L_{\alpha} = L_{\alpha}(q,\dot{q},\ddot{q}). 
\end{equation}
The extreme values of the local action integral defined by $L_\alpha$ are computed through its variation:
\begin{equation}\label{serd-00}
\delta \int_a^bL_{\alpha}(q,\dot{q},\ddot{q} )dt=0. 
\end{equation}
In this picture, we have the local second order Euler-Lagrange equations: 
\begin{equation}\label{local2EL}
\frac{\partial L_{\alpha} }{\partial q^i} - \frac{d }{dt} \frac{\partial L_{\alpha} }{\partial \dot{q}^i}  +  \frac{d^2 }{dt^2} \frac{\partial L_{\alpha}}{\partial \ddot{q}^i}  =0.
\end{equation}
Assume that on the (non-trivial) intersection of two local charts, say  
 $T^2V_\alpha \cap T^2V_\beta$, the two local Lagrangians are related as follows
\begin{equation}\label{laglog++--}
e^{\sigma_\alpha(q)}L_\alpha (q,\dot{q},\ddot{q} )  =e^{\sigma_\beta(q)}L_\beta (q,\dot{q},\ddot{q} ).
\end{equation}

The locally conformal character in \eqref{laglog++--} allows us neither to glue the set $\{L_\alpha\}$ of local second order Lagrangians nor the local Euler-Lagrange equations \eqref{local2EL} in global pictures. To define a global picture, we introduce the following local second order Lagrangian function:
\begin{equation}\label{serd-11}
L|_{\alpha}(q,\dot{q},\ddot{q} ) = e^{\sigma_{\alpha}(q)} L_{\alpha}(q,\dot{q},\ddot{q}  ).
\end{equation}
In light of \eqref{laglog++--}, the local Lagrangian functions defined through \eqref{serd-11} can be glued up to a global Lagrangian function, which we denote by $L$, on $T^2Q$. That, we now have a global Lagrangian function $L$ with local realization $L|_{\alpha}$ on $T^2V_\alpha$. We apply variational analysis to arrive at the equations of motion governed by this Lagrangian $L$. To this end, we replace  
 $L_\alpha$ in the action \eqref{serd-00} with 
 $L$ then we compute 
\begin{equation}
 \begin{split}
& \delta \int_a^b e^{-\sigma_{\alpha}}L 
dt = \int_a^b \Big(-e^{-\sigma_{\alpha}}\frac{\partial \sigma_\alpha}{\partial q^i}L  \delta q^i + e^{-\sigma_{\alpha}}\frac{\partial L  }{\partial q^i}\delta q^i \\ & \quad  \qquad 
 \qquad\qquad\qquad \qquad + e^{-\sigma_{\alpha}}\frac{\partial L }{\partial \dot{q}^i}\delta \dot{q}^i   
 +   e^{-\sigma_{\alpha}}\frac{\partial L  }{\partial \ddot{q}^i}\delta \ddot{q}^i\Big)dt
 \end{split}
\end{equation}
for fixed endpoints. By applying the method of integration by parts iteratively, and omitting the boundary terms, one arrives at the following:
\begin{equation}
 \begin{split}
 \delta \int_a^b e^{-\sigma_{\alpha}}L  dt& = \int_a^b \Big(-e^{-\sigma_{\alpha}}\frac{\partial \sigma_\alpha}{\partial q^i}L  + e^{-\sigma_{\alpha}}\frac{\partial L }{\partial q^i} \\ & \quad  \qquad - \frac{d}{dt}\big(e^{-\sigma_{\alpha}}\frac{\partial L  }{\partial \dot{q}^i}\big)  
  +  \frac{ d^2 }{dt^2}\big(e^{-\sigma_{\alpha}}\frac{\partial L  }{\partial \ddot{q}^i}\big)  \Big)\delta {q}^i dt.
 \end{split}    
\end{equation}
We assume the arbitrariness of $ \delta q^i $ and we take the variation to be zero, the result reads
\begin{equation} \label{2LCEL}
 e^{-\sigma_{\alpha}} \frac{\partial L}{\partial q^i} - \frac{d}{dt} \left( e^{-\sigma_{\alpha}} \frac{\partial L}{\partial \dot{q}^i} \right) + \frac{d^2}{dt^2} \left( e^{-\sigma_{\alpha}} \frac{\partial L}{\partial \ddot{q}^i} \right) - e^{-\sigma_{\alpha}} \frac{\partial \sigma_\alpha}{\partial q^i} L = 0.
\end{equation}
We expand each term in \eqref{2LCEL} using the Leibniz rule, and then arrive at a more explicit formulation of the locally conformal second order Euler-Lagrange equations. We record this in the following proposition which is the first novel result in this work.
\begin{proposition}\label{prop-2}
    The locally conformal Lagrangian dynamics generated by a second order Lagrangian function $L$ with the set of conformal factors $\{\sigma_\alpha\}$ is 
\begin{equation} \label{2nd-ord-direct-}
  \frac{\partial L }{\partial q^i}- \frac{d}{dt}   \frac{\partial L }{\partial \dot{q}^i} + \frac{ d^2 }{dt^2}   \frac{\partial L }{\partial \ddot{q}^i}      = \mathcal{A}^2_i[L]
  \end{equation}
where the quantity on the right-hand side is given by
\begin{equation}\label{A-2}
  \mathcal{A}^2_i[L]=\varphi_{i}L  - \varphi_{k}\dot{q}^{k} \frac{\partial L } {\partial {\dot{q}}^i} + \Big(  \varphi_{k}\ddot{q}^{k}   + \varphi_{kl}\dot{q}^{k}\dot{q}^{l}  + 2\varphi_{k}\dot{q}^{k}\frac{d}{dt} -  \varphi_{k}\varphi_{l}\dot{q}^{k}\dot{q}^{l}\Big)\frac{\partial L } {\partial {\ddot{q}}^i},
\end{equation}
with the following notation:
 \begin{equation}
\varphi_{k} =\frac{\partial \sigma_\alpha}{\partial q^k},\qquad \varphi_{kl} =\frac{\partial^2 \sigma_\alpha}{\partial q^k \partial q^l}. 
 \end{equation}
\end{proposition}

Notice that if the conformal relation is trivial, then we have $ \mathcal{A}^2_i(L) = 0$, hence \eqref{2nd-ord-direct-} turns out to be the second order Euler-Lagrange equations \eqref{2EL}. On the other hand, if the Lagrangian $ L$ does not depend on the acceleration, then the third term in \eqref{2nd-ord-direct-} drops, and the quantities $ \mathcal{A}^2_i[L]$ in \eqref{A-2} reduce to $ \mathcal{A}_i[L]$ in \eqref{A-1}, and thus the locally conformal second order Euler-Lagrange equations \eqref{2nd-ord-direct-} becomes the locally conformal Euler-Lagrange equations \eqref{LCEL} of the first order.

\section{Locally Conformal Higher Order  Lagrangian Functions} \label{sec:-Bell}
 
 In this and the following two  sections, we generalize the locally conformal second order Lagrangian theory developed in the previous section to the $n$-th order. To this end, at first, we   consider the local charts $\{V_\alpha\}$ in \eqref{chart} for a manifold $Q$ then assume the induced local chart for the  $n$-th order tangent bundle $T^nQ$ as follows
\begin{equation}
 T^nQ=\bigsqcup_\alpha T^nV_\alpha = \bigsqcup_\alpha V_\alpha \times \mathbb{R}^r \times \mathbb{R}^r \times \dots \times\mathbb{R}^r.
\end{equation}
 Notice that if  $Q$  is an $r$ dimensional manifold, then 
$T^nQ$  an $(n+1)r$-dimensional.  

Assume a local $n$-th order Lagrangian function that depends up to $n$-th time derivative of position $q$: 
\begin{equation}
L_{\alpha} = L_{\alpha}(q,\dot{q},\ddot{q}, \dots, q_{(n)}).
\end{equation}
The extreme values of the local action are computed through the variation as follows:
\begin{equation}\label{serd-0} \delta \int_a^b L_{\alpha}(q, \dot{q}, \ddot{q}, \ldots, q_{(n)}) dt = 0. 
\end{equation} 
This yields the local Euler-Lagrange equations for $L_{\alpha}$ as 
\begin{equation}\label{HOEL}
 \sum_{k=0}^{n} (-1)^k \frac{d^k}{dt^k}   \frac{\partial L_{\alpha} }{\partial {q}^i_{(k)} }       =0.
\end{equation}

We further assume that on the intersections of the local charts, the local Lagrangian functions do not coincide but are related by conformal factors. For example, on $T^nV_\alpha \cap T^nV_\beta$, we have that
\begin{equation}\label{laglog++}
e^{\sigma_\alpha(q)}L_\alpha (q,\dot{q},\ddot{q}, \dots, q_{(n)})  =e^{\sigma_\beta(q)}L_\beta (q,\dot{q},\ddot{q}, \dots, q_{(n)}).
\end{equation}
The conformal relations on local Lagrangian functions, given in \eqref{laglog++}, do not allow us to glue local higher order Lagrangians into a global one. Accordingly, we define the following local $n$-th order Lagrangian function:
\begin{equation}\label{serd-1}
L|_{\alpha}(q,\dot{q},\ddot{q}, \dots, q_{(n)}) = e^{\sigma_{\alpha}(q)} L_{\alpha}(q,\dot{q},\ddot{q}, \dots, q_{(n)}).
\end{equation}
See that, the local Lagrangian functions defined through \eqref{serd-1} can be glued up to a global Lagrangian function $L$ on $T^nQ$. In this case, the local picture of the global Lagrangian function $L$ on $T^nQ$  is $L|_{\alpha}$. To get the equation of motion for the Lagrangian function $L$, we substitute the local Lagrangian functions $L_\alpha$ in the action \eqref{serd-0} with $L$ in \eqref{serd-1}. Accordingly, we have 
\begin{equation}
 \begin{split}
& \delta \int_a^b e^{-\sigma_{\alpha}}L dt = \int_a^b \Big(-e^{-\sigma_{\alpha}}\frac{\partial \sigma_\alpha}{\partial q^i}L \delta q^i + e^{-\sigma_{\alpha}}\frac{\partial L }{\partial q^i}\delta q^i \\ & \quad  \qquad 
 \qquad\qquad\qquad \qquad + e^{-\sigma_{\alpha}}\frac{\partial L }{\partial \dot{q}^i}\delta \dot{q}^i   
 + \dots + e^{-\sigma_{\alpha}}\frac{\partial L }{\partial {q}^i_{(n)}}\delta {q}_{(n)}^i\Big)dt
 \end{split}
\end{equation}
for fixed endpoints. After applying the method of integration by parts, then omitting the boundary terms, the variation turns out to be
\begin{equation}\label{n-variation}
 \begin{split}
& \delta \int_a^b e^{-\sigma_{\alpha}}L dt = \int_a^b \Big(-e^{-\sigma_{\alpha}}\frac{\partial \sigma_\alpha}{\partial q^i}L + e^{-\sigma_{\alpha}}\frac{\partial L }{\partial q^i} \\ & \quad  \qquad - \frac{d}{dt}\big(e^{-\sigma_{\alpha}}\frac{\partial L }{\partial \dot{q}^i}\big)  
  +   \dots + (-1)^n \frac{ d^n}{dt^n} \big(e^{-\sigma_{\alpha}}\frac{\partial L }{\partial {q}^i_{(n)}}\big)\Big)\delta {q}^i dt.
 \end{split}    
\end{equation}
We set the variation \eqref{n-variation} to zero and get 
\begin{equation} \label{nLCEL}
 \sum_{s=0}^{n} (-1)^s \frac{d^s}{dt^s} \Big(e^{-\sigma_{\alpha}} \frac{\partial L  }{\partial {q}^i_{(s)}}  \Big) - e^{-\sigma_{\alpha}} \frac{\partial \sigma_\alpha}{\partial q^i} L  = 0 . 
\end{equation}
To obtain a more explicit expression for the dynamical equations \eqref{nLCEL}, we expand each term individually. This requires a substantial study of the Leibniz and chain rules in combinatorial notation, which will be addressed in the upcoming section.

\section{Some Combinatorial Notations and Bell's Polynomials} \label{sec:-Bell-}
This section examines the terms in \eqref{nLCEL}. This leads us to get rid of the exponential functions and permits us to write the dynamical equations in a more compact form. The combinatorial properties of the chain rule for composition of functions are studied in \cite{comb-derivatives}.

We begin with the $k$-th time derivative  term (for $n\geq k>1$) in \eqref{nLCEL} as follows
\begin{equation}\label{hoo-1}
 \begin{split}
 \frac{ d^s}{dt^s} \Big(e^{-\sigma_{\alpha}} \frac{\partial L } {\partial {q}^i_{(s)}} \Big)&  = \sum_{a=0}^{s}  \binom{s}{a} \frac{ d^{s-a}}{dt^{s-a}} (e^{-\sigma_{\alpha}}) \frac{ d^{a}}{dt^{a}}\frac{\partial L } {\partial {q}^i_{(s)}} .
 \end{split}   
\end{equation} 
Observe that, in \eqref{hoo-1}, there appear the higher order derivatives of the conformal factor $ e^{-\sigma_{\alpha}} $. A straightforward but tedious calculation using repeated applications of the chain rule yields
\begin{equation} \label{bell}
  \frac{ d^{s}}{dt^{s}} (e^{-\sigma_{\alpha}}) = e^{-\sigma_{\alpha}} 
\mathcal{B}_{s},  
\end{equation}
where the second term on the right-hand side is 
\begin{equation} \label{bell01}
\mathcal{B}_{s}=\sum_{m=1}^{s} \Big[ \Phi_m \cdot B_{s,m}(\dot{q},\ddot{q}, \dots, q_{(s)})    \Big].
\end{equation}
Let us precisely define the terms in \eqref{bell01} one by one.

The first term $\Phi_m$ in   \eqref{bell01} refers to the  combinatorial product  
\begin{equation} \label{Phi}
\Phi_m :=\sum_{\mathfrak{p} \in P} (-1)^{|\mathfrak{p}|}\displaystyle \prod_{S \in \mathfrak{p}}  \frac{\partial^{|S|}\sigma_{\alpha} }{\prod_{u \in S}\partial q^u}
\end{equation}
determining the partial derivatives of the conformal factor $\sigma_{\alpha}$. 
To define $ \Phi_m $ more explicitly and show its relation to $ m $, we consider the index set $ M = \{i_1,i_2,\dots,i_m\} $. The set $ P $ consists of all partitions of $ M $. We denote by $ \mathfrak{p} $ a partition in $ P $ and by $ S $ a block of the partition $\mathfrak{p} $. Here, $ |S| $ denotes the number of elements in the block $ S $, while $ |\mathfrak{p}| $ denotes the number of blocks in the partition $\mathfrak{p} $. 

\begin{example}

In the first-order theory, where $ n=1 $, we have $ k=1 $ in \eqref{nLCEL}, and $ m $ is also equal to $ 1 $.
In this case, the partition set is a singleton, i.e., $ P = \{i\} $, hence $ M = \{i\} $. Therefore, we have only one block, and the combinatorial notation is
\begin{equation}\label{Phi-1}
 \Phi_1=\sum_{\mathfrak{p} \in P} (-1)^{|\mathfrak{p} |}\displaystyle \prod_{S \in \mathfrak{p} }  \frac{\partial^{|S|}\sigma_{\alpha} }{\prod_{u \in S}\partial q^u}=
\sum_{p \in \{i\}} (-1)^{1}\displaystyle  \frac{\partial\sigma_{\alpha}}{\partial q^{(p)}} = -   \frac{\partial\sigma_{\alpha}}{\partial q^i} = - \varphi_i,
\end{equation}
where $\partial\sigma_{\alpha}/ \partial q^i= \varphi_i$. 
\end{example}

\begin{example}
For the second order theory ($ n=2 $), we have $ k=1 $ and $ k=2 $. The former case was examined in the previous paragraph, so we now study $ k=2 $. In this case,  $ P = \{i,j\} $. The set $ P $ has two partitions: $ \big\{ \{i\}, \{j\} \big\} $ and $ \big\{ \{i,j\} \big\} $.  
We get:
\begin{equation}\label{Phi-2}
\begin{split}
\Phi_2 &=\sum_{\mathfrak{p}  \in P} (-1)^{|\mathfrak{p}|}\displaystyle \prod_{S \in \mathfrak{p}}  \frac{\partial^{|S|}\sigma_{\alpha} }{\prod_{u \in S}\partial q^u} \\ &= 
 (-1)^{\vert\{ \{i\},\{j\} \}\vert}\displaystyle \prod_{S \in \{ \{i\},\{j\} \}}  \frac{\partial^{|S|}\sigma_{\alpha} }{\prod_{u \in S}\partial q^u} 
   + 
 (-1)^{\vert\{ \{i,j\} \}\vert}\displaystyle \prod_{S \in \{ \{i,j\} \}}  \frac{\partial^{|S|}\sigma_{\alpha} }{\prod_{u \in S}\partial q^u} 
 \\ &  = \frac{\partial^{|\{i\}|}\sigma_{\alpha} }{\prod_{u \in \{i\}}\partial q^u} \frac{\partial^{|\{j\}|}\sigma_{\alpha} }{\prod_{u \in \{j\}}\partial q^u} -  \frac{\partial^{|\{ \{i\},\{j\} \}|}\sigma_{\alpha} }{\prod_{u \in \{ \{i\},\{j\} \}}\partial q^u} 
 \\ &
 = \frac{\partial\sigma_{\alpha} }{\partial q^i}\frac{\partial\sigma_{\alpha} }{\partial q^j} -  \frac{\partial^{2}\sigma_{\alpha} }{ \partial q^i \partial q^j}=\varphi_i \varphi_j - \varphi_{ij} ,
 \end{split}
\end{equation}
where $\partial^2\sigma_{\alpha}/ \partial q^i \partial q^j= \varphi_{ij}$. 
\end{example}
\begin{example}
For the third-order theory ($ n=3 $), $ k $ runs from $ 1 $ to $ 3 $. The cases $ k=1 $ and $ k=2 $ are the same as those computed in the previous examples. Now, we take $ k=3 $. In this case, we get the set $ M = \{i,j,k\} $, which has five partitions:
\begin{equation}
\begin{split}
    \big\{ \{i\},\{j\},\{k\}\big\}, ~  \big\{ \{i\},\{j,k\}\big\},~ \big\{ \{j\},\{k,i\}\big\}, ~ \big\{ \{k\},\{i,j\}\big\}, ~ \big\{ \{i,j,k\} \big\}. 
\end{split}
\end{equation}
Therefore, $\Phi_3$ has five terms calculated as follows:
\begin{equation*}
  \begin{split}
&\Phi_3 =\sum_{\mathfrak{p}  \in P} (-1)^{|\mathfrak{p}|}\displaystyle \prod_{S \in \mathfrak{p}}  \frac{\partial^{|S|}\sigma_{\alpha} }{\prod_{u \in S}\partial q^u}=
 (-1)^{|\{ \{i\},\{j\},\{k\}\}|}\displaystyle \prod_{S \in \{ \{i\},\{j\},\{k\}\}}  \frac{\partial^{|S|}\sigma_{\alpha} }{\prod_{u \in S}\partial q^u}
 \\ &\quad   + (-1)^{| \{i\},\{j,k\}\}|}\displaystyle \prod_{S \in  \{\{i\},\{j,k\}\}}  \frac{\partial^{|S|}\sigma_{\alpha} }{\prod_{u \in S}\partial q^u} + (-1)^{| \{ \{j\},\{i,k\}\}|}\displaystyle \prod_{S \in  \{ \{j\},\{i,k\}\}}  \frac{\partial^{|S|}\sigma_{\alpha} }{\prod_{u \in S}\partial q^u}
 \\ &\quad + (-1)^{| \{ \{k\},\{i,j\}\}|}\displaystyle \prod_{S \in \{\{ \{k\},\{i,j\}\}}  \frac{\partial^{|S|}\sigma_{\alpha} }{\prod_{u \in S}\partial q^u} + (-1)^{| \{\{i,j,k\}\}|}\displaystyle \prod_{S \in \{\{i,j,k\}\}}  \frac{\partial^{|S|}\sigma_{\alpha} }{\prod_{u \in S}\partial q^u}.
  \end{split}
\end{equation*}
We count he number of elements in the partition, and run the multiplication in the denominators accourding to the partitions. These lead us to the following calculation: 
\begin{equation} \label{Phi-3}
  \begin{split}
\Phi_3 &= -  \frac{\partial^{\vert\{i\}\vert}\sigma_{\alpha} }{\prod_{u \in \{i\}}\partial q^u}  \frac{\partial^{\vert\{j\}\vert}\sigma_{\alpha} }{\prod_{u \in \{j\}}\partial q^u}  \frac{\partial^{\vert\{k\}\vert}\sigma_{\alpha} }{\prod_{u \in \{k\}}\partial q^u} +\frac{\partial^{\vert\{i\}\vert}\sigma_{\alpha} }{\prod_{u \in \{i\}}\partial q^u}\frac{\partial^{\vert\{j,k\}\vert}\sigma_{\alpha} }{\prod_{u \in \{j,k\}\ }\partial q^u}
  \\ &\qquad  +\frac{\partial^{\vert\{j\}\vert}\sigma_{\alpha} }{\prod_{u \in \{j\}}\partial q^u}\frac{\partial^{\vert\{i,k\}\vert}\sigma_{\alpha} }{\prod_{u \in \{i,k\}\ }\partial q^u} +\frac{\partial^{\vert\{k\}\vert}\sigma_{\alpha} }{\prod_{u \in \{k\}}\partial q^u}\frac{\partial^{\vert\{i,j\}\vert}\sigma_{\alpha} }{\prod_{u \in \{i,j\}\ }\partial q^u}\\&  \qquad  - \frac{\partial^{|\{\{i,j,k\}\}|}\sigma_{\alpha} }{\prod_{\{\{i,j,k\}\} \in S}\partial q^u}\\& =
      -\frac{\partial \sigma_{\alpha}} {\partial {q}^i} \frac{\partial \sigma_{\alpha}} {\partial {q}^j} \frac{\partial \sigma_{\alpha}} {\partial {q}^k} + \frac{\partial \sigma_{\alpha}} {\partial {q}^i} \frac{\partial^2 \sigma_{\alpha}} {\partial {q}^j \partial {q}^k} + \frac{\partial \sigma_{\alpha}} {\partial {q}^j}\frac{\partial^2 \sigma_{\alpha}} {\partial {q}^k \partial {q}^i}   + \frac{\partial \sigma_{\alpha}} {\partial {q}^k} \frac{\partial^2 \sigma_{\alpha}} {\partial {q}^i \partial {q}^j} \\& \qquad - \frac{\partial^3 \sigma_{\alpha}} {\partial {q}^i \partial {q}^j \partial {q}^k} 
      \\& =-\varphi_{i}\varphi_{j}\varphi_{k}+ \varphi_{i} \varphi_{jk} +  \varphi_{j} \varphi_{ki} +\varphi_{k} \varphi_{ij} -  \varphi_{ijk},
  \end{split}
\end{equation}
where we take  $ \partial^3\sigma_{\alpha}/{\partial q^i}{\partial q^j}{\partial q^j}= \varphi_{ijk} $.
\end{example}

We have defined the combinatorial quantities $\Phi_m$ in \eqref{bell01} and have examined particular cases $m=1,2$ and $3$. Now, we define the second term $B_{s,m}$ in the notation \eqref{bell01}. They are the partial exponential Bell polynomials \cite{Bell} with its variables being the time derivatives of $q$ \cite{advcombinatorics}: 
\begin{equation} \label{expbell}
  B_{s,m}(\dot{q}, \ddot{q}, \ldots, q_{(s-m+1)}) = \sum \frac{s!}{c_1! c_2! \ldots c_s!} \frac{\dot{q}^{c_1} \ddot{q}^{c_2} \ldots {q}^{c_s}_{(s)}}{(1!)^{c_1} (2!)^{c_2}\ldots(s!)^{c_s}} ,
\end{equation}
where the sum is taken over all non-negative integers $c_{i}$ such that the following two equations
\begin{equation}
    \begin{split}
       & c_1 +  2c_2 + \cdots + sc_s = s, \\
       &  c_1 +  c_2 + \cdots + c_s = m \\
    \end{split}
\end{equation}
hold.  In Equation \eqref{expbell}, $\dot{q}^{c_1}$ denotes the $c_1$'th power of $\dot{q}$. Its meaning in index notation  will be clear in the examples given below.

Let us present some examples of these polynomials and calculate $\mathcal{B}_a$ terms. These will be useful for the corollaries that will be recorded in the upcoming section. 
\begin{example}
If $v=1$ for $B_{v,m}$, then $m$ can only take the value $1$. In this case, we have the Bell's polynomial $B_{1,1}(\dot{q}) = \dot{q}^i$. Then, to multiply the Bell's polynomial $B_{1,1}$ with the combinatorial quantity $\Phi_1$ in \eqref{Phi-1} leads us to arrive at $\mathcal{B}_{1}$ appearing in \eqref{bell01}:
\begin{equation}\label{B-1}
    \begin{split}
    \mathcal{B}_{1}= 
    B_{1,1}(\dot{q}) \cdot \Phi_1 = - \frac{\partial\sigma_{\alpha}}{\partial q^i}\dot{q}^i = -\varphi_{i}\dot{q}^i.    
    \end{split}
\end{equation} 
\end{example}

\begin{example}
If $v=2$ for $B_{v,m}$ then $m$ can take the values $1$ and $2$. These give the following Bell's polynomials 
\begin{equation}
\begin{split} 
&B_{2,1}(\dot{q},\ddot{q}) = \ddot{q}^i,\quad
B_{2,2}(\dot{q},\ddot{q}) = \dot{q}^i\dot{q}^j. 
\end{split}
\end{equation}
Multiplying the Bell's polynomials with the combinatorial quantities $\Phi_1$ in \eqref{Phi-1} and $\Phi_2$ in \eqref{Phi-2}, we determine $\mathcal{B}_{2}$ (in \eqref{bell01}) as
\begin{equation} \label{B-2}
    \begin{split}
     \mathcal{B}_{2}& = \sum_{m=1}^{2} \Big[ B_{2,m}(\dot{q},\ddot{q}) \cdot \Phi_m  \Big]
     =B_{2,1}(\dot{q},\ddot{q}) \cdot \Phi_1 + 
     B_{2,2}(\dot{q},\ddot{q})\cdot \Phi_2   \\
     & = - \frac{\partial\sigma_{\alpha}}{\partial q^i} \ddot{q}^i + \Big(\frac{\partial\sigma_{\alpha} }{\partial q^i}\frac{\partial\sigma_{\alpha} }{\partial q^j} -  \frac{\partial^{2}\sigma_{\alpha} }{ \partial q^i \partial q^j}\Big) \dot{q}^i\dot{q}^j = -\varphi_{i}\ddot{q}^i +  (\varphi_{i} \varphi_{j} - \varphi_{ij}  ) \dot{q}^i\dot{q}^j.
    \end{split}
\end{equation} 
\end{example}

\begin{example}
For $v=3$ for $B_{v,m}$, $m$ can take three values, namely $1$, $2$ and $3$. Hence, we have three polynomials:
\begin{equation}  B_{3,1}(\dot{q},\ddot{q},q_{(3)}) = q_{(3)}^i, \quad
B_{3,2}(\dot{q},\ddot{q},q_{(3)}) =3\ddot{q}^i\dot{q}^j,\quad
B_{3,3}(\dot{q},\ddot{q},q_{(3)}) = \dot{q}^i\dot{q}^j\dot{q}^k. 
\end{equation}
By multiplying the Bell's polynomials with the combinatorial quantities $\Phi_1$ in \eqref{Phi-1}, $\Phi_2$ in \eqref{Phi-2}, and $\Phi_3$ in \eqref{Phi-3}, we compute $\mathcal{B}_{3}$ in \eqref{bell01} as
\begin{equation} \label{B-3}
    \begin{split}
         \mathcal{B}_{3} & = \sum_{m=1}^{3} \Big[ B_{3,m}(\dot{q},\ddot{q}, q_{(3)}) \cdot \Phi_m \Big] \\
         & = B_{3,1}(\dot{q},\ddot{q},q_{(3)}) \cdot \Phi_1 + B_{3,2}(\dot{q},\ddot{q},q_{(3)}) \cdot \Phi_2 + 
     B_{3,3}(\dot{q},\ddot{q},q_{(3)}) \cdot \Phi_3 \\
     & =- \varphi_{i}q_{(3)}^i + 3\Big(\varphi_{i} \varphi_{j} - \varphi_{ij} \Big)\ddot{q}^i\dot{q}^j \\&\qquad   + \big(-\varphi_{i}\varphi_{j}\varphi_{k} + \varphi_{i}\varphi_{jk} + \varphi_{j}\varphi_{ki} +\varphi_{k}\varphi_{ij} - \varphi_{ikj} \big)\dot{q}^i\dot{q}^j\dot{q}^k.
    \end{split}
\end{equation} 
\end{example}
After examining  the tensorial quantities and the Bell's polynomials, we are ready to arrive at the locally conformal $n$-th order Euler-Lagrange equations.

\section{Locally Conformal Higher Order Euler-Lagrange Equations}\label{sec:LCnEL}

In this section, we start with the dynamical equations in \eqref{nLCEL} then substitute the combinatorial notation and Bell's polynomials into them. This gives our main result, locally conformal higher order Lagrangian dynamics. 

Accordingly, we consider \eqref{nLCEL} and write it as  
\begin{equation}
 \sum_{s=0}^{n} (-1)^s \frac{d^s}{dt^s} \Big(e^{-\sigma_{\alpha}} \frac{\partial L  }{\partial {q}^i_{(s)}}  \Big) - e^{-\sigma_{\alpha}(q)} \varphi_i L  = 0.
 \end{equation}
 For the $s$-th derivative terms, we substitute the binomial expansion given in \eqref{hoo-1}, hence arrive at 
\begin{equation} \label{nLCEL-2} 
   \sum_{s=0}^{n} (-1)^s \sum_{a=0}^{s}  \binom{s}{a} \frac{ d^{s-a}}{dt^{s-a}} (e^{-\sigma_{\alpha}}) \frac{ d^{a}}{dt^{a}}\frac{\partial L } {\partial {q}^i_{(s)}}  - e^{-\sigma_{\alpha}(q)}  \varphi_i L  = 0. 
   \end{equation}
   Then for the derivatives of $e^{-\sigma_\alpha}$, we substitute \eqref{bell}. So, we get the following expression
    \begin{equation}
     \sum_{s=0}^{n} (-1)^s \sum_{a=0}^{s}  \binom{s}{a}  e^{-\sigma_{\alpha}} 
\mathcal{B}_{s-a}  \frac{ d^{a}}{dt^{a}}\frac{\partial L } {\partial {q}^i_{(s)}}  - e^{-\sigma_{\alpha}}  \varphi_i L  = 0 .  \end{equation}
It is now possible to cancel out the exponential terms $e^{-\sigma_{\alpha}}$. Later, we collect the terms involving the partial derivatives to the right hand side as 
\begin{equation}
     \sum_{s=0}^{n} (-1)^s \frac{d^s}{dt^s}   \frac{\partial L }{\partial {q}^i_{(s)} } = \varphi_{i}L+\sum_{s=1}^{n} (-1)^{s+1} \sum_{a=0}^{s-1}\binom{s}{a} \mathcal{B}_{s-a} \frac{d^{a}}{dt^{a}} \frac{\partial L} {\partial {q}^i_{(s)}}.
\end{equation}
Thus, we arrive at the following proposition, which encapsulates the primary goal of this work: locally conformal 
$n$-th order Euler-Lagrange equations.

\begin{proposition}\label{prop-main}
     The locally conformal Lagrangian dynamics generated by an $n$-th order Lagrangian function $L$ with the set of conformal factors $\{\sigma_\alpha\}$ is 
\begin{equation} \label{2nd-ord-direct--}
 \sum_{s=0}^{n} (-1)^s \frac{d^s}{dt^s}   \frac{\partial L }{\partial {q}^i_{(s)} }         = \mathcal{A}^n_i[L]
  \end{equation}
where the quantity on the right-hand side is given by
\begin{equation}\label{A-2-}
  \mathcal{A}^n_i[L]=\varphi_{i}L+\sum_{s=1}^{n} (-1)^{s+1} \sum_{a=0}^{s-1}\binom{s}{a} \mathcal{B}_{s-a} \frac{d^{a}}{dt^{a}} \frac{\partial L} {\partial {q}^i_{(s)}}.
\end{equation}
Here, $\{\mathcal{B}_{s-a}\}$'s are the quantities in \eqref{bell01} and $
\varphi_{i} = \partial \sigma_\alpha / \partial q^i.  $
\end{proposition}
We refer to the system \eqref{2nd-ord-direct--} as the locally conformal 
$n$-th order Euler-Lagrange equations.
Let us examine Propositon \ref{prop-main} for the case $n=1$, $n=2$ and $n=3$, one by one. We start with the case $n=1$. 
\begin{example}\label{2nd-order Lag}
 For $n=1$, we have locally conformal Euler-Lagrange equations
 (of the first order) as 
 \begin{equation}\label{erer}
 \sum_{s=0}^{1} (-1)^s \frac{d^s}{dt^s}   \frac{\partial L }{\partial {q}^i_{(s)} }         = \mathcal{A}^1_i[L]
 \end{equation}
 where 
  \begin{equation}
\mathcal{A}^1_i[L]= \varphi_{i}L+  \mathcal{B}_{1} \frac{\partial L} {\partial {\dot{q}}^i}.
   \end{equation}
Here, $\mathcal{B}_{1}$ is in \eqref{B-1}. 
The dynamics in \eqref{erer} is exactly as described in Section \ref{LCEL-sec} and explicitly stated in Proposition \ref{prop-1}.
\end{example}

\begin{example}\label{2nd-order Lag+}
For $n=2$, we have locally conformal second order Euler-Lagrange equations
 \begin{equation}\label{erer-2}
 \sum_{s=0}^{2} (-1)^s \frac{d^s}{dt^s}   \frac{\partial L }{\partial {q}^i_{(s)} }         = \mathcal{A}^2_i[L]
 \end{equation}
 where 
  \begin{equation}
\mathcal{A}^2_i[L]=  \varphi_i L +\mathcal{B}_{1} \frac{\partial L} {\partial {\dot{q}}^i} - \Big(  \mathcal{B}_{2} + 2\mathcal{B}_{1} \frac{d }{dt } \Big)\frac{\partial L} {\partial {\ddot{q}}^i} .
   \end{equation}
Here $\mathcal{B}_{1}$ is in \eqref{B-1},
   and $\mathcal{B}_{2}$ is in \eqref{B-2}. 
   The dynamics \eqref{erer-2} is 
   is identical to that outlined in Proposition \ref{prop-2} of Section \ref{sec:2ndLC}.
   \end{example}

   \begin{example} For $n=3$, we have locally conformal third order Euler-Lagrange equations: 
 \begin{equation}
  \sum_{s=0}^{3} (-1)^s \frac{d^s}{dt^s}   \frac{\partial L }{\partial {q}^i_{(s)} }         = \mathcal{A}^3_i[L].
  \end{equation} 
where 
\begin{equation} \label{3rd-ord-example}
    \mathcal{A}^3_i[L]=  \varphi_i L +\mathcal{B}_{1} \frac{\partial L} {\partial {\dot{q}}^i} - \Big(  \mathcal{B}_{2} + 2\mathcal{B}_{1} \frac{d }{dt } \Big)\frac{\partial L} {\partial {\ddot{q}}^i} + \Big(  \mathcal{B}_{3} + 3\mathcal{B}_{2} \frac{d }{dt } + 3\mathcal{B}_{1} \frac{d^2 }{dt^2 } \Big)\frac{\partial L} {\partial {q}^i_{(3)}}.
\end{equation}
 Here $\mathcal{B}_{1}$ is in \eqref{B-1},
 $\mathcal{B}_{2}$ is in \eqref{B-2}, amd $\mathcal{B}_{3}$ is in \eqref{B-3}. 


\end{example}

\section{A Locally Conformal Chiral Oscillator} \label{sec:examp}

As the base manifold, consider two dimensional punctured Euclidean space $Q=\mathbf{R}^2-\{0\}$ where $0$ stands for the origin, then present the closed one-form 
\begin{equation} \label{Lee-ex}
\vartheta=2\frac{xdy-ydx}{x^2+y^2}
\end{equation}
which fails to be exact on the whole $Q$. After a cut in $Q$, we determine polar coordinates $x=r\cos\phi$, $y=r\sin\phi$. In this local chart, the closed one-form $\vartheta$ becomes the exact one-form $2d\phi$, so $\sigma_\alpha=2\phi$.

The Lagrangian function for the chiral oscillator \cite{MR1479488} in Cartesian coordinates is
\begin{equation} \label{chiral-osc}
    L(q^i, \dot{q}^i, \ddot{q}^i) = - \frac{\lambda}{2} \epsilon_{ij} \dot{q}^i \ddot{q}^j + \frac{m}{2} \dot{q}_{i} \dot{q}^i
\end{equation}
Here, $\lambda$ and $m$ are nonzero constants, $q^i = (x,y)$ and $\epsilon_{ij}$ is the skew-symmetric Levi-Civita symbol with $\epsilon_{12}=1$ such that $i,j$ runs from $1$ to $2$.
Euler-Lagrange equations for the chiral oscillator are 
\begin{equation}
\lambda \epsilon_{ij} {q}^j_{(3)} - m \ddot{q}_{i} = 0. 
\end{equation}
We cite \cite{ccaugatay2019total} for the Dirac analysis of this formalism. 

In the locally conformal picture, we take the following Lagrangian function:
\begin{equation}
    L_{\alpha}(q^i, \dot{q}^i, \ddot{q}^i) = - \frac{\lambda}{2} e^{-\sigma_{\alpha}(q)}\epsilon_{ij} \dot{q}^i \ddot{q}^j + \frac{m}{2} e^{-\sigma_{\alpha}(q)} \dot{q}_{i} \dot{q}^i
\end{equation}
Then we plug this local Lagrangian into Equation \eqref{local2EL}. The resulting expansion is of the form in \eqref{2LCEL}. Direct calculation gives the following locally conformal Euler-Lagrange equations
\begin{equation}
    \begin{split}
    & \lambda \epsilon_{ij} {q}^j_{(3)} - m \ddot{q}_{i} = \varphi_{i}(- \frac{\lambda}{2} \epsilon_{kj} \dot{q}^k \ddot{q}^j + \frac{m}{2} \dot{q}_{k} \dot{q}^k) - \varphi_{l}\dot{q}^l (- \frac{\lambda}{2} \epsilon_{ij} \ddot{q}^j + \frac{m}{2} \dot{q}_{i}) \\
    &   + (\varphi_{l} \ddot{q}^l + \varphi_{lm} \dot{q}^l \dot{q}^m - \varphi_{l} \varphi_{m} \dot{q}^l\dot{q}^m)\frac{\lambda}{2}\epsilon_{ij}\dot{q}^j + \lambda \varphi_{l}\dot{q}^l \epsilon_{ij}\ddot{q}^j .
    \end{split}
\end{equation}
Notice that this has exactly the same form with Equation \eqref{A-2} if we set 
\begin{equation}
\frac{\partial L}{\partial \dot{q}^i} = - \frac{\lambda}{2} \epsilon_{ij} \ddot{q}^j + m \dot{q}_{i},\qquad \frac{\partial L}{\partial \ddot{q}^i} = \frac{\lambda}{2} \epsilon_{ij}\dot{q}^j. 
 \end{equation} 

\section{Conclusion and Future Work}

 In this work, we have explored the locally conformal analysis of higher order Lagrangian dynamics. Following a brief review of the fundamental structures of locally conformally symplectic (LCS) manifolds, we revisited the locally conformal Euler-Lagrange equations, as outlined in Proposition \ref{prop-1}. We then reviewed higher order Lagrangian dynamics and presented, as the first novel result, the locally conformal second order Euler-Lagrange equations in Proposition \ref{prop-2}. Subsequently, we extended the analysis to an arbitrary 
$n$-th order, deriving the locally conformal 
$n$-th order Euler-Lagrange equations in Proposition \ref{prop-main}. 

We list two open questions for further investigation:
\begin{itemize}

  \item LCS manifolds are specific examples of Jacobi manifolds \cite{Lichnerowicz-Jacobi}. More precisely, a Jacobi manifold is integrable, with its even-dimensional leaves being LCS manifolds and its odd-dimensional leaves being contact manifolds. A geometric Hamilton-Jacobi theory for contact dynamics has been studied in the continuous framework \cite{de2021hamilton} and the discrete framework \cite{math12152342}. Similarly, a geometric Hamilton-Jacobi theory for LCS dynamics is presented in \cite{EsenLeonSarZaj1}, and for higher order Lagrangian dynamics in \cite{EsenLeonSar2}. We plan to develop a geometric Hamilton-Jacobi theory for the locally conformal higher order Euler-Lagrange equations derived herein.

    \item Another approach for discussing locally conformal higher order Lagrangian dynamics is  discrete framework. The discretization of Lagrangian dynamics (of the first order) is available in the literature, see \cite{MarsWest}. We refer to \cite{BobeSuri99,MaPeSh99} for discrete Lagrangian dynamics from a reduction perspective, and to \cite{EsSuDiscrete,MaDeMa2006} for discrete dynamics within the Lie groupoid setting.  The study of discrete Lagrangian dynamics for higher order Lagrangian dynamics is given in \cite{MR2232878}. Additionally, the discretization of locally conformal Lagrangian dynamics in the first-order case is established in \cite{MR4747740}. In an upcoming work, we aim to study the locally conformal higher order Euler-Lagrange equations within a discrete framework.
\end{itemize}

\bibliographystyle{abbrv}
\bibliography{references}{}

\end{document}

%% file: preamble.tex
\usepackage[a4paper,margin=23mm,top=30mm]{geometry}

\usepackage{amsfonts, amsmath, amssymb, amsgen, amsthm, amscd}
\usepackage{newtxtext,newtxmath}
\usepackage[utf8]{inputenc} 

\usepackage{color}

\usepackage[all]{xy}

\usepackage{hyperref}
\usepackage{mathtools,slashed}
\usepackage{setspace}
\usepackage{enumerate}

\usepackage{enumerate}

\renewcommand{\geq}{\geqslant}

\numberwithin{equation}{section}

\newtheorem{theorem}{Theorem}[section]
\newtheorem{proposition}{Proposition}[section]

\theoremstyle{definition}

\newtheorem{example}[theorem]{Example}

